\documentstyle [12pt,epsf]{article}

\makeatletter
\def\maketitle{\par
 \begingroup
 \def\thefootnote{\fnsymbol{footnote}}
 \def\@makefnmark{\mbox{$^\@thefnmark$}}
 \@maketitle
 \@thanks
 \endgroup
 \setcounter{footnote}{0}
 \let\maketitle\relax
 \let\@maketitle\relax
 \gdef\@thanks{}\gdef\@author{}\gdef\@title{}\let\thanks\relax}
\def\@maketitle{\vspace*{0.9cm}
{\hsize\textwidth
 \linewidth\hsize \centering
 {\normalsize \bf \@title \par} \vskip 0.3cm  {\normalsize  \@author \par}}}

\def\thefootnote{\mbox{\noindent$\fnsymbol{footnote}$}}
    \long\def\@makefntext#1{\noindent$^{\@thefnmark}$#1}

\def\section{\@startsection{section}{1}{\z@}{1.5ex plus 0.5ex minus
   1.2ex}{1.3ex plus .1ex}{\normalsize\bf}}
\def\subsection{\@startsection{subsection}{2}{\z@}{1.5ex plus 0.5ex minus
    1.2ex}{1.3ex plus .1ex}{\normalsize\em}}

\def\@sect#1#2#3#4#5#6[#7]#8{\ifnum #2>\c@secnumdepth
     \def\@svsec{}\else
     \refstepcounter{#1}\edef\@svsec{\ifnum #2=1 \@sectname\fi
	\csname the#1\endcsname.\hskip 1em }\fi
     \@tempskipa #5\relax
      \ifdim \@tempskipa>\z@
	\begingroup #6\relax
	  \@hangfrom{\hskip #3\relax\@svsec}{\interlinepenalty \@M #8\par}
	\endgroup
       \csname #1mark\endcsname{#7}\addcontentsline
	 {toc}{#1}{\ifnum #2>\c@secnumdepth \else
		      \protect\numberline{\csname the#1\endcsname}\fi
		    #7}\else
	\def\@svsechd{#6\hskip #3\@svsec #8\csname #1mark\endcsname
		      {#7}\addcontentsline
			   {toc}{#1}{\ifnum #2>\c@secnumdepth \else
			     \protect\numberline{\csname the#1\endcsname}\fi
		       #7}}\fi
     \@xsect{#5}}

\def\@sectname{}

\def\thebibliography#1{\section*{{{\normalsize
\bf References }
\rule{0pt}{0pt}}\@mkboth
  {REFERENCES}{REFERENCES}}\list
  {{\arabic{enumi}.}}{\settowidth\labelwidth{{#1}}%
    \leftmargin\labelwidth  \frenchspacing
    \advance\leftmargin\labelsep
    \itemsep=-0.2cm
    \usecounter{enumi}}
    \def\newblock{\hskip .11em plus .33em minus -.07em}
    \sloppy
    \sfcode`\.=1000\relax}

\def\@cite#1#2{\unskip\nobreak\relax
    \def\@tempa{$\m@th^{\hbox{\the\scriptfont0 #1}}$}%
    \futurelet\@tempc\@citexx}
\def\@citexx{\ifx.\@tempc\let\@tempd=\@citepunct\else
    \ifx,\@tempc\let\@tempd=\@citepunct\else
    \let\@tempd=\@tempa\fi\fi\@tempd}
\def\@citepunct{\@tempc\edef\@sf{\spacefactor=\the\spacefactor\relax}\@tempa
    \@sf\@gobble}

\def\citenum#1{{\def\@cite##1##2{##1}\cite{#1}}}
\def\citea#1{\@cite{#1}{}}

\newcount\@tempcntc
\def\@citex[#1]#2{\if@filesw\immediate\write\@auxout{\string\citation{#2}}\fi
  \@tempcnta\z@\@tempcntb\m@ne\def\@citea{}\@cite{\@for\@citeb:=#2\do
    {\@ifundefined
       {b@\@citeb}{\@citeo\@tempcntb\m@ne\@citea\def\@citea{,}{\bf ?}\@warning
       {Citation `\@citeb' on page \thepage \space undefined}}%
    {\setbox\z@\hbox{\global\@tempcntc0\csname b@\@citeb\endcsname\relax}%
     \ifnum\@tempcntc=\z@ \@citeo\@tempcntb\m@ne
       \@citea\def\@citea{,}\hbox{\csname b@\@citeb\endcsname}%
     \else
      \advance\@tempcntb\@ne
      \ifnum\@tempcntb=\@tempcntc
      \else\advance\@tempcntb\m@ne\@citeo
      \@tempcnta\@tempcntc\@tempcntb\@tempcntc\fi\fi}}\@citeo}{#1}}
\def\@citeo{\ifnum\@tempcnta>\@tempcntb\else\@citea\def\@citea{,}%
  \ifnum\@tempcnta=\@tempcntb\the\@tempcnta\else
   {\advance\@tempcnta\@ne\ifnum\@tempcnta=\@tempcntb \else \def\@citea{--}\fi
    \advance\@tempcnta\m@ne\the\@tempcnta\@citea\the\@tempcntb}\fi\fi}

\def\abstract{\if@twocolumn
\section*{Abstract}	    
\else \small
\begin{center}
{ABSTRACT\vspace{-.5em}\vspace{0pt}}
\end{center}
\quotation
\fi}
\def\endabstract{\if@twocolumn\else\endquotation\fi}

\def\fnum@figure{Fig. \thefigure}

\long\def\@makecaption#1#2{
   \vskip 10pt
   \setbox\@tempboxa\hbox{\small #1. #2}
   \ifdim \wd\@tempboxa >\hsize	   
      \small #1. #2\par		   
   \else			   
      \hbox to\hsize{\hfil\box\@tempboxa\hfil}
   \fi}

\def\eps{\epsilon}
\def\gw{g_{\rm w}}
\def\alphaw{\alpha_{\rm w}}
\def\Mw{M_{\rm w}}
\def\tr{{\rm tr}}
\def\Tc{T_{\rm c}}

\def\gsim{ \,\, \vcenter{\hbox{$\buildrel{\displaystyle >}\over\sim$}} \,\,}

\begin {document}

\begin {flushright}
UW-PT-94-13 \\
October 1994
\end {flushright}

\title{THE ELECTROWEAK PHASE TRANSITION, PART 1\\
Review of Perturbative Methods%
\footnote{
   Talk presented at the conference Quarks `94: Vladimir, Russia, 1994.
   This work was supported by the U.S. Department of Energy,
   Grant No.\ DE-FG06-91ER40614.
}%
}
\author{
  Peter Arnold\\
  {\em Dept. of Physics, FM-15, Univ. of Washington, Seattle, WA 98115, USA}
}
\maketitle
\setlength{\baselineskip}{14pt}

\vspace{0.7in}

The goal of this talk, and the following one by Larry Yaffe, will be to
investigate the order and strength of the electroweak phase transition.
I will review what can be done with standard perturbative methods and
how such methods sometimes break down in cases of interest.  In part 2,
Larry Yaffe will discuss the application of $\eps$ expansion techniques
to study those cases where standard perturbative methods fail.

The reason for studying the electroweak phase transition is that it
plays a crucial role in scenarios of electroweak baryogenesis.
Recall Sakharov's three conditions for any scenario of baryogenesis:
(1) baryon number violation, (2) C and CP violation, and (3) disequilibrium.
As I shall briefly review, standard electroweak theory provides
the required violation of baryon number.  The standard model also
provides C and CP violation, though the strength of such violation
may not be sufficient to generate the observed baryon excess unless the
Higgs sector is non-minimal---an issue discussed in other talks at this
conference.  As I shall discuss later, the role of the electroweak phase
transition is to provide the required disequilibrium, and its success in
this role depends on the order and strength of the transition.

In our talks, Larry and I will stick to a simple toy model: the minimal
standard model with a single doublet Higgs.  I call this a toy model because
one probably needs to incorporate extra Higgs bosons into the theory to
have sufficient CP violation for electroweak baryogenesis.  But multiple
Higgs models have all sorts of unknown parameters in them, which makes it
difficult to plot results in any simple way.  It makes sense to first
refine one's techniques for studying the phase transition in the simpler
one-Higgs model.  With a little bit of work, everything we do should be
straightforwardly extendable to more complicated models.  For simplicity,
we shall also generally ignore $\sin^2\theta_{\rm w}$ and focus on pure
SU(2) gauge-Higgs theory.


\section{Lightning review of electroweak B violation}

I shall take a moment to quickly review baryon number (B) violation in
standard electroweak theory;%
\footnote{
   For a sample of various reviews of the subject, try
   ref.~\protect\citenum{b violation}.
   For a review of electroweak baryogenesis, try
   ref.~\protect\citenum{baryogenesis}.
}
the formula for the B violation rate
will later be relevant to motivating some of the important issues concerning
the electroweak phase transition.

\begin {figure}
\vbox
    {%
    \begin {center}
	\leavevmode
	
	\epsfbox [150 300 500 500] {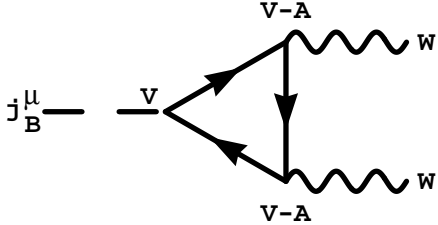}
    \end {center}
    \caption
	{%
	\label {anomaly}
        The triangle anomaly for the baryon number current in electroweak
        theory.
	}%
    }%
\end {figure}

Baryon number violation is a bit strange in standard electroweak theory
because it can't happen perturbatively.  All of the vertices in a Feynman
diagram conserve quark number; whenever a quark line enters a diagram,
it remains a quark line and must eventually leave again, thus conserving
baryon number.  However, baryon number is violated {\it non}-perturbatively
due to the electroweak anomaly shown in fig.~\ref{anomaly}.  This anomaly is
closely analogous to the usual axial anomaly of QCD or QED.  In the
electroweak case, however, the
axial nature of the anomaly appears in the gauge couplings rather than
in the current.  The formula for the anomaly is the same as
in the QCD case,
\begin {equation}
   \partial_\mu j^\mu \sim \gw^2 F \tilde F \,,
\end {equation}
except that the field strengths $F$ are for the weak SU(2) fields rather
than the gluon fields.  Integrating both sides gives a formula for the
amount of baryon number violation in any process:
\begin {equation}
   \Delta B \sim \gw^2 \int d^4x \, F \tilde F \,.
\label {del B eq}
\end {equation}
\begin {figure}
\vbox
    {%
    \begin {center}
	\leavevmode
	
	\epsfbox [130 320 480 480] {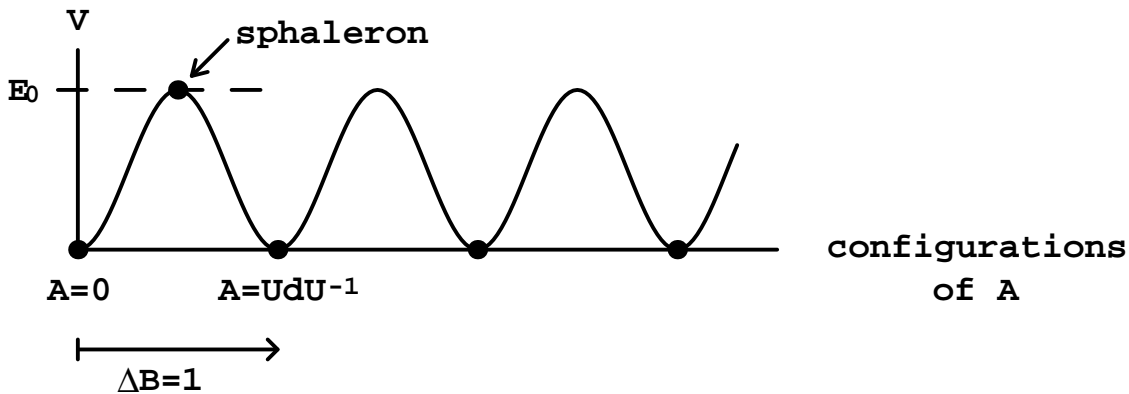}
    \end {center}
    \caption
	{%
	\label {sphaleron}
        Qualitative picture of gauge configuration space for a B violating
        transition.
	}%
    }%
\end {figure}
Note that, in order to get a $\Delta B$ of order 1, the field strengths
$F$ must be of order $1/\gw$.  So any process which violates baryon number
involves large, non-perturbative excursions away from the vacuum $F=0$.
Also note that large field strengths imply large energies, and so any
transition with $\Delta B \sim 1$ requires passing
through intermediate gauge field configurations with non-negligible energy.
This situation is depicted schematically in fig.~\ref{sphaleron}.
The horizontal
axes denotes the sequence of gauge field configurations a particular
process passes through when violating baryon number via
(\ref{del B eq}); the vertical axis denotes the potential energy of
those configurations.  $E_0$ is the potential energy barrier separating
the initial gauge vacuum $A=0$ from the final gauge vacuum, which is just
a gauge transform.  The configuration corresponding to the minimum
potential energy barrier for this process is known as the sphaleron.

At zero energy, the only way to get from one vacuum to the next, and so
produce B violation through the anomaly, is by quantum tunneling.
Because the barrier is non-perturbatively large, the probability for such
tunneling is exponentially suppressed and turns out to be
\begin {equation}
   \hbox{rate} \sim e^{-4\pi/\alphaw} \sim 10^{-170} = \hbox{zero} \,.
\end {equation}
Imagine instead the early universe at temperatures large compared to the
barrier energy $E_0$.  Such a hot thermal bath will excite states with
energies large compared to $E_0$, and these can cross the barrier classically
rather than tunneling beneath it.  The B violation rate will {\it not} be
exponentially suppressed.  Now consider an intermediate situation where the
universe is hot but the temperature is smaller than $E_0$.  Then there's
still some chance that a random thermal fluctuation will have enough energy
to cross the barrier, and this probability is naively given by a simple
Boltzmann factor:
\begin {equation}
   \hbox{rate} \sim e^{-\beta E_0} \sim e^{-\beta \Mw/\alphaw} \,,
\label {B rate}
\end {equation}
where $\beta$ is the inverse temperature.  The estimate of $E_0$ above
may be understood from (1) the earlier observation that field strengths must be
order $1/\gw$, which means energies are $1/\gw^2$, and (2) the fact that
$E_0$ has dimensions of mass and $\Mw$ is the natural mass scale of
electroweak theory:
\begin {equation}
   E_0 \sim \Mw/\alphaw \sim \hbox{a few TeV} \,.
\end {equation}
All of this can be made more rigorous, and the numerical coefficients
in these equations can be deduced, but the simple parameter dependence
I am exhibiting here will be all that I'll use for this talk.
Note that my general sloppiness in writing equations extends even to
exponents: the last exponent in (\ref{B rate}) has some numerical
coefficient in it which I haven't bothered to show.


\section {Disequilibrium}

In GUT scenarios for baryogenesis, all the relevant physics occurs at
temperatures of order $10^{16}$ GeV and the expansion of the universe
directly provides the disequilibrium needed for baryogenesis.
In electroweak scenarios for baryogenesis, the relevant physics occurs
at temperatures of order the weak scale, when the universe is very much
older, expanding very much more slowly, and so very much closer to
equilibrium.  Back of the envelope estimates show that the expansion is
then far too slow to produce the observed number of baryons.

But there is other physics that is taking place around the same time---namely,
the electroweak phase transition---and this transition can potentially supply
the needed element of disequilibrium.  If the transition is second order,
the universe never departs significantly from equilibrium during the
transition.  However, if it is first order (and
sufficiently strongly so), then it proceeds by the nucleation, expansion,
and percolation of bubbles of the new phase inside the old---a non-equilibrium
process.  Each point in space will feel a non-equilibrium jolt as a bubble
wall sweeps by converting it from the symmetric phase ($\phi{=}0$) to
the asymmetric phase ($\phi\not=0$), and so baryogenesis in these scenarios
takes place on (or near) the bubble walls.  Back of the envelope estimates
have shown that, for some models of the Higgs sector, one has a first-order
phase transition and can get in the
ballpark of postdicting the observed baryon-to-photon ratio
$n_{\rm B}/s \sim 10^{-10}$.

To sharpen these back of the envelope estimates, there are many interesting
but complicated problems one could study.  How do you accurately compute
the bubble wall profile?\ the bubble wall velocity?\ the amount of baryogenesis
as the wall sweeps by?  All of these non-equilibrium problems are complicated,
and so I'm going to focus instead on a simpler problem relevant to the
success or failure of electroweak baryogenesis scenarios.


\section{A simpler constraint on models}

After the phase transition is completed, and the universe settles down into
the new, asymmetric phase with $\phi\not=0$, it had better be the case
that baryon number violation is turned off.  Otherwise, the universe will
simply relax back to its preferred equilibrium state of $B=0$ and all of
the baryogenesis that occured during the phase transition will be washed
away.  To turn off B violation, we need the rate $e^{-\beta E_0}$ to
be small compared to the expansion rate of the universe, which means
the exponent
\begin {equation}
   \beta E_0 \sim \Mw/\alphaw T \sim \gw\phi/\alphaw T
\end {equation}
\begin {figure}
\vbox
    {%
    \begin {center}
	\leavevmode
	
	\epsfbox [140 300 500 550] {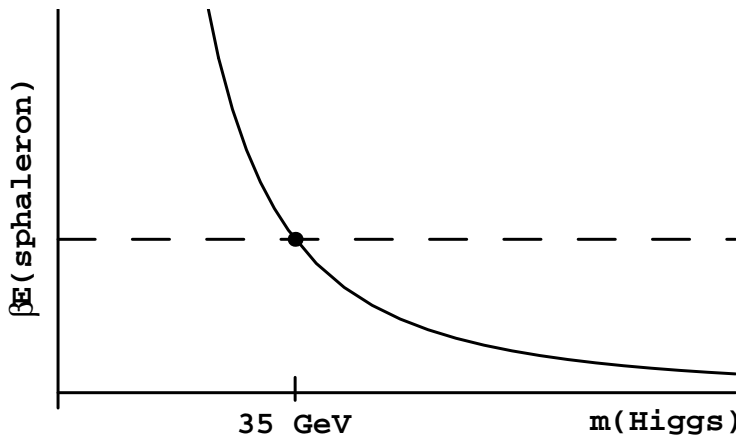}
    \end {center}
    \caption
	{%
	\label {figB}
        The Boltzmann exponent for baryon number violation vs.\ the
        zero-temperature Higgs mass.
	}%
    }%
\end {figure}%
must be {\it large} in the asymmetric phase just after the phase transition
is completed.  In the minimal standard model, a leading-order calculation
of this exponent (which I will review in a moment) depends qualitatively
on the zero-temperature Higgs boson mass as shown in fig.~\ref{figB}.
I will explain why the exponent depends inversely on the Higgs mass, but
for the moment let's consider the consequences.
A comparison of the B violation rate to the expansion rate of the universe
was made by Shaposhnikov\cite{Shaposhnikov}
and later improved by Dine {\it et al.}\cite{Dine}
Using a leading-order perturbative calculation,%
\footnote{
   ``Leading'' order here means leading order after improvement by
   resumming hard thermal loops (daisies).
}
the requirement that B
violation be turned off
puts a lower bound on the exponent and hence an upper bound on the Higgs
mass, as depicted in the figure.
For the minimal standard model, this bound on the Higgs mass is
roughly 35--40 GeV, which is ruled out by the experimental lower bound
of 60 GeV.
So minimal standard model baryogenesis appears to fail.
If one makes the Higgs sector more complicated, it is possible to evade
these bounds, which is yet another reason to study multiple Higgs models.

However, the situation is more complicated than I have just made it out
to be.  As I shall discuss, the {\it leading}-order calculation used to
derive these constraints may be inadequate and higher-order corrections
may be crucial.  But first, let me outline how the
leading-order calculation is made.


\section{The leading-order calculation}

Consider the classical Higgs potential:
\begin {equation}
   V_0 \sim -\mu^2\phi^2 + \lambda\phi^4 \,.
\label {V0 eq}
\end {equation}
$V_0$ basically tells us the ``vacuum'' energy as a function of $\phi$,
and at zero temperature the ground-state is determined by minimizing it.
At finite temperature, the ground state is determined by minimizing the
free energy.  At finite temperature, the system is not in vacuum---there
is a plasma of real, on-shell particles such as W's, quarks, and leptons,
and all contribute to the free energy.  For the sake of pedagogy, let me
just focus on the contribution of W's in the plasma, and for the time
being let's ignore interactions.  The free energy of an ideal
Bose gas is something you can easily look up in a graduate textbook:
\begin {equation}
   \Delta F \sim T \int d^3 k \, \ln\left(1 - e^{-\beta E_k}\right) \,,
\end {equation}
where the relativistic energy is just
\begin {equation}
   E_k = \sqrt{\vec k^2 + \Mw^2} \sim \sqrt{\vec K^2 + g^2\phi^2} \,.
\end {equation}
Note that the W mass is proportional to $g\phi$ in a background Higgs
field $\phi$, and so the W gas contribution $\Delta F$ to the free energy
is a function of $\phi$.  At high temperature ($T >> \Mw$), $\Delta F$
can be expanded in powers of $1/T$ to give
\begin {equation}
   \Delta F \sim \# T^4 + \# \Mw^2 T^2 - \# \Mw^3 T + \cdots \,,
\label{W gas}
\end {equation}
where I haven't bothered showing the numerical coefficients \#.
Henceforth I won't bother writing in the \# signs either.

\begin {figure}
\vbox
    {%
    \begin {center}
	\leavevmode
	
	\epsfbox [150 260 500 530] {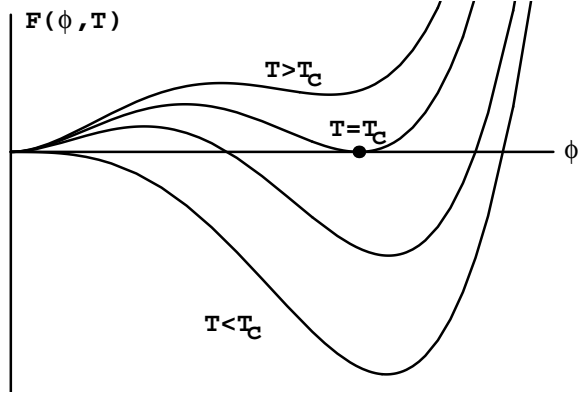}
    \end {center}
    \caption
	{%
	\label {Vfig}
        The form of the free energy, as a function of $\phi$, for
        different temperatures.
	}%
    }%
\end {figure}

To get the total free energy, we just add the ``vacuum'' contribution
(\ref{V0 eq}) and the W gas contribution (\ref{W gas}):
\begin {equation}
   F \sim V_0 + \Delta F \sim
   \hbox{const.} + (-\mu^2+g^2 T^2)\phi^2 - g^3 \phi^3 T + \lambda\phi^4
   \cdots \,.
\label {F eq}
\end {equation}
The $\phi$-independent term ``const.''\ doesn't affect the properties of
the transition and will be ignored.  The $g^2 T^2 \phi^2$ term comes from
the $\Mw^2 T^2$ term of (\ref{W gas}) and is responsible for driving the
phase transition: it turns the curvature of the free energy at $\phi=0$ from
concave-down at $T=0$ to concave-up at sufficiently large $T$.
The cubic term $-g^3 \phi^3 T$ comes from the $-\Mw^3 T$ term and is
responsible for making the phase transition first-order, as depicted in
fig.~\ref{Vfig}, rather than second-order.

We are now in a position to estimate the order of magnitude (or more
specifically, the parameter dependence) of quantities related to the phase
transition.  Examine fig.~\ref{Vfig} and consider the critical temperature at
which the two ground states are degenerate, and consider the region of
$\phi$ which, roughly, encompasses the maximum and asymmetric minimum
of $F(\phi)$.
The only way for the free energy to have that shape is if the
quadratic, cubic, and quartic terms of (\ref{F eq}) all have the same
order of magnitude in that region of $\phi$.  If the quadratic term
were negligible, it wouldn't curve up at the origin; if the cubic term
were negligible, it wouldn't curve down later; and if the quartic term
were negligible, it wouldn't be turning up again.  So
\begin {equation}
   (-\mu^2 + g^2 T^2)\phi^2 \sim g^3 \phi^3 T \sim \lambda \phi^4 \,.
\end {equation}
The last relation, in particular, then easily yields
\begin {equation}
   \phi \sim {g^3\over\lambda} T \,.
\end {equation}

Now let's return to the rate of B violation in the asymmetric phase.
The Boltzmann exponent
$\beta E_0$ is then
\begin {equation}
   {\Mw\over g^2 T} \sim {\phi\over gT}
   \sim {g^2\over\lambda}
   \sim {M^2({\rm W})_{T=0} \over m^2({\rm Higgs})_{T=0}} \,.
\label{B rate exponent}
\end {equation}
This is how one finds the inverse dependence on the Higgs mass that
was depicted in fig.~\ref{figB}.


\section{Review of finite-temperature formalism}

Our next goal will be to discuss the validity of the above leading-order
treatment, where the W bosons were treated as an ideal gas.
Discussing corrections due to interactions requires a little more careful
treatment of finite temperature calculations, and so in this section I
shall very briefly review the formalism of finite-temperature field theory.
Recall that the basic tool for studying equilibrium questions at finite
temperature is the partition function:
\begin {equation}
   Z = \tr \, e^{-\beta H} \,.
\end {equation}
A path integral expression for the partition function may be easily found
by noting that $e^{-\beta H}$ is nothing but the time-evolution operator for
an imaginary amount of time $t = i\beta$.  So, in the exact same way one
derives the path integral for infinite-time evolution to attack
zero-temperature problems, one may derive the completely analogous result
\begin {equation}
   Z = \int [{\cal D}\phi] \exp\left[ -\int\nolimits_0^\beta d\tau
         \int d^3 x {\cal L}_{\rm E}(\phi) \right] \,,
\end {equation}
where ${\cal L}_{\rm E}$ is the Euclidean action density.
The only difference is that the integral in the exponent is over Euclidean
time $\beta$ rather than over infinite real time.  Also, the trace in
the partition function is implemented by requiring the ``initial'' state
be the same as the ``final'' state, which amounts to requiring the boundary
condition
\begin {equation}
   \phi(\tau{=}0, \vec x) = \phi(\tau{=}\beta, \vec x)
\end {equation}
on the path integral.

Since the only difference between finite temperature and zero temperature is
the extent of Euclidean time, the only difference in Feynman
rules will be that, when integrating over internal momenta $k$, only the
frequencies $k_0 = 2\pi n T$ which are periodic in time $\beta$ are relevant.
So Fourier integrals get replaced by Fourier sums, and the only difference
in the Euclidean Feynman rules is the replacement
\begin {equation}
   \int d^4 k \to T \sum_{k_0} \int d^3 k \,,
   \qquad
   k_0 = 2\pi n T \,,
\label {momentum integral}
\end {equation}
for integrations over internal momenta.  Note for later the factor of
$T$ in front of the Fourier series sum, which makes the dimensions the
same as $d^4 k$.

Now consider what happens in the large temperature limit $\beta\to 0$.
In this case, the extent of the (Euclidean) temporal dimension shrinks to
zero.  So the four-dimensional theory reduces to an effective
three-dimensional theory of the static ($k_0=0$) modes.
(More precisely, this reduction takes place if we are studying equilibrium
quantities at distance scales large compared to $\beta$.)
I should note that fermions turn out to have {\it anti}-periodic boundary
conditions and so cannot have any static modes.
As a result, fermions completely decouple from long-distance, equilibrium
physics in the high-temperature limit.

The introduction of temperatures into Feynman rules via
(\ref{momentum integral}) may seem a bit formal.  However, it turns out
to have a fairly simple physical interpretation when one does actual
calculations.  If one carries out the Euclidean frequency sum for a
simple one-loop integral, one finds%
\footnote{
  For a review, try ref.~\citenum{Kapusta}.
}
\begin {equation}
\hbox{
  \setlength{\unitlength}{0.12in}
  \begin {picture}(45,10)
    \thicklines
    \put(2,5){\circle{4}}
    \put(2,3){\line( 1,-1){2}}
    \put(2,3){\line(-1,-1){2}}
    \put(3,7.6){\vector(-1,0){2}}
    \put(2,8){$k$}
    \put(8.5,4.7){=}
    \put(16,5){\circle{4}}
    \put(16,3){\line( 1,-1){2}}
    \put(16,3){\line(-1,-1){2}}
    \put(14.7,4.7){$T{=}0$}
    \put(17,7.6){\vector(-1,0){2}}
    \put(16,8){$k$}
    \put(20.5,4.7){+}
    \put(24,4){$\displaystyle{
           \int {d^3 k\over (2\pi)^3 2E_k} \, {1 \over e^{\beta E_k} - 1}
    }$}
    \put(40,4){\line( 1,-1){2}}
    \put(40,4){\line(-1,-1){2}}
    \put(40,4){\line(-3, 2){3.5}}
    \put(40,4){\line( 3, 2){3.5}}
    \put(37,6.8){\vector(3,-2){2}}
    \put(41,5.4){\vector(3, 2){2}}
    \put(36,6.8){$k$}
    \put(43.4,6.8){$k$}
    \put(45,3){.}
  \end {picture}
}
\label {finite T display}
\end {equation}
The first term on the right-hand side denotes the zero-temperature result.
The second term---which contains all the temperature dependence---is nothing
more than the amplitude for the external particle to forward scatter off of
a real, physical particle present in the thermal bath:
the $1/(e^{\beta E_k}-1)$ is the Bose probability for finding such a particle,
and the $d^3 k/(2\pi)^3 2E_k$ is just the usual measure for phase space.


\section{Loop Expansion Parameter}
\label{loop expansion section}

We're now in a position to discuss when leading-order calculations
are adequate at finite temperature.  The basic cost of adding a loop
to a diagram at high temperature is\footnote{
   More precisely, this is the cost once one has absorbed hard thermal
   loops (daisies) into propagators, which is something I'll briefly
   discuss later.
}
\begin {equation}
   {g^2 T\over \hbox{physics scale}} \sim {g^2 T \over \Mw} \,.
\label {loop parameter}
\end {equation}
The $g^2$ is just the usual cost of extra coupling constants.
The factor of $T$ is the explicit factor of $T$ from the Fourier
sum (\ref{momentum integral}) associated with the additional loop
momentum.
But the cost of adding a loop should be dimensionless,
so $g^2 T$ must be divided by whatever gives the mass scale of the
problem---in this case, $\Mw$.  Because of the factor of $T/\Mw$,
the loop expansion is not necessarily small at high temperature!
The criteria that the loop expansion parameter be small, and that
therefore a perturbative expansion around mean field theory be
useful, is known to condensed matter physicists as the
{\it Ginzburg} criteria.

The loop expansion parameter (\ref{loop parameter}) is very important,
so let's understand it in several different ways.
First, consider adding a loop with internal momentum $k$ to any
diagram, denoted by a shaded circle below.  As discussed earlier,
the effect of finite temperature on loops is to incorporate the
physics of particles forward-scattering off of real particles in the plasma:
\def\blob{
  \begin{picture}(2,2)(0.35,0)
    \thicklines
    \put(0,0){\circle{2}}
    \thinlines
    \put(-0.7,-0.7){\line(1,1){1.3}}
    \put(-0.9,0.0){\line( 1, 1){1.0}}
    \put( 0.9,0.0){\line(-1,-1){1.0}}
    \put(-0.85,-0.4){\line( 1, 1){1.2}}
    \put( 0.85, 0.4){\line(-1,-1){1.2}}
  \end{picture}
}
\begin {equation}
\setlength{\unitlength}{0.15in}
\begin {picture}(40,15)
  \thicklines
  \put(5,11){\oval(4,2.5)[t]}
  \put(4,11){\oval(2,2.5)[bl]}
  \put(6,11){\oval(2,2.5)[br]}
  \put(5,9.5){\blob}
  \put(6,12.8){\vector(-1,0){2}}
  \put(5,13.2){$k$}
  \put(9.5,10.2){$\sim$}
  \put(13,10){$\displaystyle{
         \int {d^3 k\over (2\pi)^3 2E_k} \, {1 \over e^{\beta E_k} - 1}
  }$}
  \put(28,10){\blob}
  \put(27.2,10.3){\line(-3,2){2.5}}
  \put(28.8,10.3){\line( 3,2){2.5}}
  \put(24.8,12.6){\vector(3,-2){2}}
  \put(29.2,11.2){\vector(3, 2){2}}
  \put(24,12.6){$k$}
  \put(31.6,12.6){$k$}
  \put(9.5,4.2){$\sim$}
  \put(8.3,3.2){{\small large T}}
  \put(13,4){$\displaystyle{
         \int {d^3 k\over (2\pi)^3 2E_k} \, ~~~{T\over E_k}
  }$}
  \put(28,4){\blob}
  \put(27.2,4.3){\line(-3,2){2.5}}
  \put(28.8,4.3){\line( 3,2){2.5}}
  \put(24.8,6.6){\vector(3,-2){2}}
  \put(29.2,5.2){\vector(3, 2){2}}
  \put(24,6.6){$k$}
  \put(31.6,6.6){$k$}
  \put(32,3){.}
\end {picture}
\end {equation}
The only temperature dependence in the last line is the explicit factor
of $T$, and so the loop expansion parameter is proportional to $g^2 T$
as before.
The rest of the expression in the last line must give something
determined by the mass scale of the problem (assuming the diagram is
sufficiently convergent in the ultraviolet) and is dominated by
$E_k \sim \Mw$.
So the origin of the large
$T/\Mw$ factor in the loop expansion parameter is simply the
divergent behavior of the Bose factor $1/(e^{\beta E_k}-1)$ as
$E_k \to 0$; there are a large number of low-energy bosons present
in a high-temperature plasma.

Let's understand the loop expansion parameter in yet another way.
As mentioned earlier, the high-temperature limit $\beta\to 0$
reduces the four-dimensional Euclidean theory to an effective
three-dimensional theory of the static ($k_0=0$) modes.
Restricting attention to the static modes, the integrand of the
path integral then has the form
\begin {equation}
   e^{-S_{\rm E}}
   = \exp\left[ -{1\over g^2} \int\nolimits_0^\beta d\tau
         \int d^3 x {\cal L}_{\rm E} \right]
   \to \exp\left[ - {1\over g^2 T} \int d^3x {\cal L}_{\rm E} \right] \,.
\label {3d reduction}
\end {equation}
In the first equality, I have normalized the fields so that the
coupling constant appears explicitly out in front of the action as
$1/g^2$.  When I specialize to static field configurations, the
Euclidean time integration becomes trivial, giving a factor of $1/T$.
But now we see that $g^2$ always appears in the combination $g^2 T$.
Then dimensional analysis gives us the loop expansion parameter
(\ref{loop parameter}) as before.

Note that for pure, unbroken gauge theory (where there is no Higgs $\phi$
to give a mass to the W),
(\ref{3d reduction}) shows that the only scale in the theory would be
$g^2 T$ itself.  If we were to study the physics at that scale,
then the loop expansion parameter given by the left-hand side of
(\ref{loop parameter}) would be order one; the physics is strongly-coupled
even though $g^2$ is small.  This is known as the infrared
(or ``magnetic mass'') problem of high-temperature non-Abelian gauge
theory.


\section{Why life is not simple}

We're now finally in a position to discuss under what conditions
perturbation theory might be adequate to study the electroweak phase
transition.  Our loop expansion parameter (\ref{loop parameter}) is
nothing other than the inverse of the B violation rate exponent
(\ref{B rate exponent}), and so we may borrow the earlier analysis
of the parameter dependence:\footnote{
  For a more detailed discussion of the loop expansion, see
  ref.~\citenum{Arnold&Espinosa}.
}
\begin {equation}
   {g^2 T\over\Mw}
   \sim {\lambda\over g^2}
   \sim {m^2({\rm Higgs})_{T=0} \over M^2({\rm W})_{T=0}} \,.
\label {loop parameter 2}
\end {equation}
A basic result of this review, which should be remembered for Larry Yaffe's
talk on the $\epsilon$ expansion, is then:
\begin {center}
  THE LOOP EXPANSION WORKS WHEN $\lambda\ll g^2$.
\end {center}
Now once can wonder how well perturbation theory is doing at the
upper bound $m$(Higgs) = 35 GeV that we earlier discussed for
electroweak baryogenesis in the minimal standard model.  Is 35 GeV
small compared to the W mass of 80 GeV?  The answer, of course, depends
on all the factors of 2 and $\pi$ that left out of the
simple minded ``$\sim$'' equalities presented in this review.
But there's a way to check it.  One can (1) formally assume
$\lambda{\ll}g^2$, (2) explicitly compute the next-to-leading order
({\it i.e.}\ two-loop) correction to the free-energy $F(\phi)$, and then
(3) see if the correction is numerically large for a 35 GeV Higgs.

I should note in passing that from (\ref{loop parameter 2}) one sees
that my constant use of the high-temperature limit $T{\gg}\Mw$ is
justified provided $g^4{\ll}\lambda$.  So the calculation actually
formally assumes
\begin {equation}
   g^4 \ll \lambda \ll g^2 \,,
\end {equation}
where the first inequality is for the high-temperature expansion and
the second for the loop expansion.

\begin {figure}
\vbox
    {%
    \begin {center}
	\leavevmode
	
	\epsfbox [150 250 500 550] {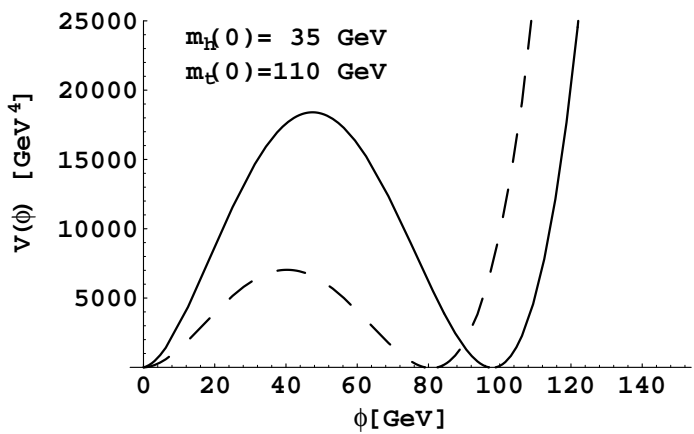}
    \end {center}
    \caption
	{%
	\label {figa}
	The effective potential at the critical temperature for
	$m_{\rm h}(0)$ = 35 GeV and $m_{\rm t}(0)$ = 110 GeV.
	The dashed and solid lines
	are the one-loop and two-loop results respectively.
        [Why a 110 GeV top mass? Because this is an old graph.
        But the results aren't particularly sensitive to $m_{\rm t}$.]
	}%
    }%
\end {figure}

The details may be found in refs.~\citenum{Arnold&Espinosa} and
\citenum{Bagnasco&Dine}, but the result of the calculation
is shown in fig.~\ref{figa}.
Both the one-loop and 2-loop results for $F(\phi)$
are shown at the temperature that makes their minima degenerate.  The
value of $\phi$ in the asymmetric ground state has only shifted by about
20\%.  On the other hand, the height of the hump has shifted by almost a
factor of three!  The moral is that for {\it some} quantities,
{(35 GeV)/(80 GeV)} is {\it not} small, and perturbation theory cannot
necessarily be trusted.%
\footnote{
  There are a variety of caveats to this conclusion.  First of all, I have
  not shown you the corrections to any {\it physical} quantity.  Beyond
  leading-order, the exact value of the height of the hump (which becomes
  complex), does not have any obvious physical interpretation, and the
  expectation of $\phi$ is gauge-dependent.  (The 2-loop result shown was
  computed in Landau gauge.)  There are plenty of examples in the world
  where perturbation theory is under reasonable numerical control for
  physical quantities but not for unphysical ones.

  ~~~~In addition, Farakos {\it et al.}\cite{Farakos} believe
  that using the renormalization
  group to improve some logarithmic corrections may bring perturbative
  calculations under numerical control, though some of those authors believe
  perturbation theory fails disastrously for different reasons, as discussed by
  Shaposhnikov in this conference.
}
In particular, we have no way to know {\it a priori}
whether the B violation rate might not be a quantity which gets
substantial corrections.  (Though the B violation rate exponent is
proportional to $\phi$ at leading order, this simple relation does not
hold beyond leading order.)


\section{More on the reduction to 3 dimensions}

When discussing the loop expansion parameter back in section
\ref{loop expansion section}, I assumed that the relevant
``physics scale'' for any diagram was determined by particle masses.
This is true if diagrams are ultraviolet (UV) convergent and so dominated
by their infrared behavior.  However, here's a quadratically divergent
diagram, for which it's not true:
\begin {equation}
   \matrix{
     \buildrel k \over \longleftarrow \cr
     \hbox{
        
        \epsfbox[150 150 500 700]{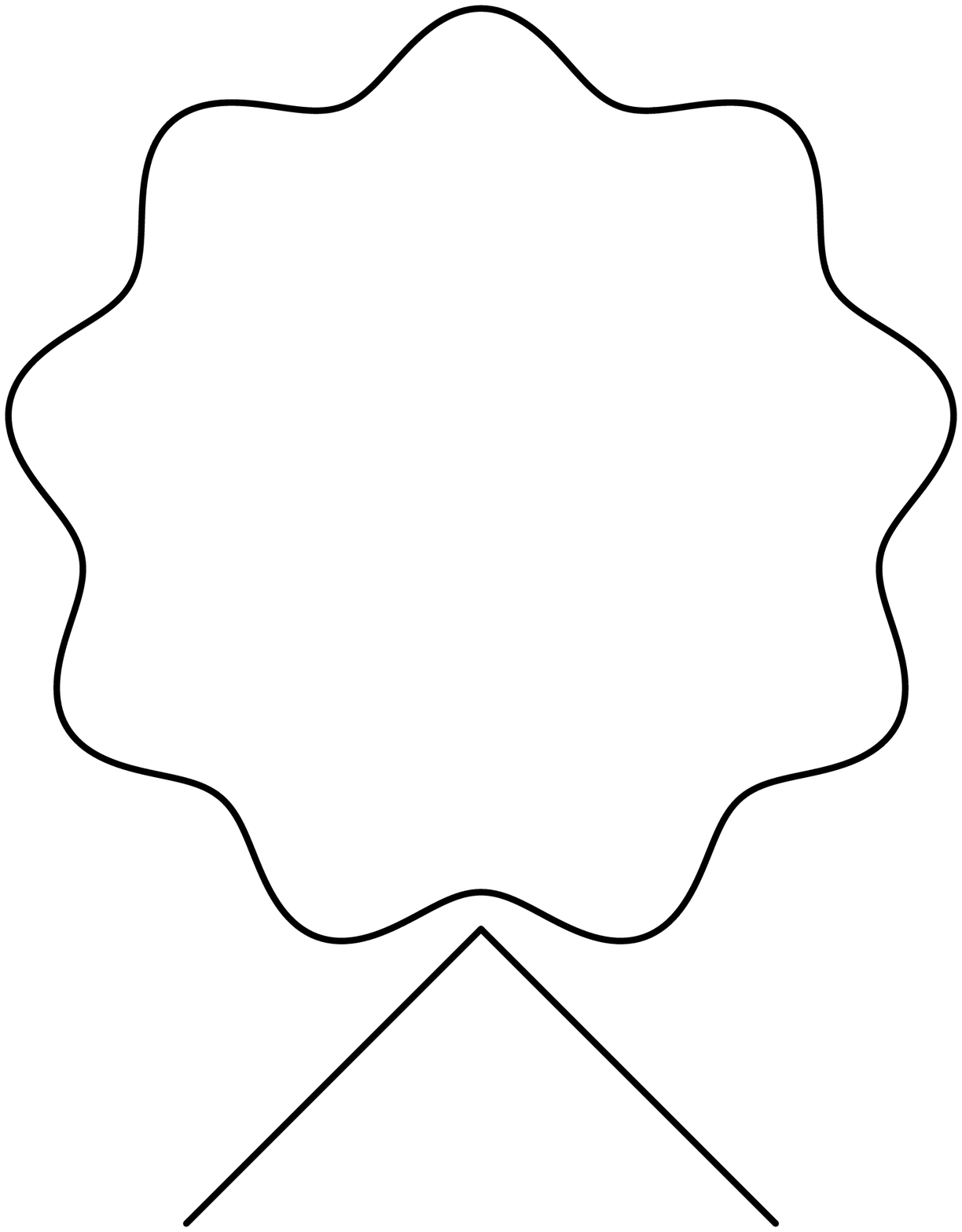}
     }\cr
   }
   \qquad = \qquad \hbox{($T{=}0$ stuff)}
   ~+~ g^2 T^2
\label {Debye mass}
\end {equation}
[Note: as usual, I'm leaving out numerical coefficients and just showing the
order of magnitude of terms.]
As in (\ref{finite T display}), the finite-temperature piece of this
diagram comes from interactions with real particles of momentum $k$ in
the plasma.  The quadratic divergence is then cut-off by $T$ for this
piece because there are no such particles with $k{\gg}T$.  This gives
the result of order $g^2 T^2$ indicated above, and the diagram is dominated
by loop momenta $k$ of order $T$.

Because the important momentum scale for this diagram is $T$ (and not the
particle masses), this diagram is sensitive to the Euclidean temporal
direction.  That is, $k_0 \not= 0$ modes are {\it not} suppressed in
UV divergent diagrams.  But this is the usual story for the decoupling of
heavy degrees of freedom in field theory.  At large distances compared
to $\beta=1/T$, the Euclidean time dimension decouples {\it except} for
renormalizations of the masses and couplings of the theory.
The $g^2 T^2$ contribution to (\ref{Debye mass}) is just the renormalization
of the scalar mass in matching the original four-dimensional theory to
the effective three-dimensional theory at long distances.

The way to systematically construct the effective three-dimensional theory,
and to relate its parameters to those of the more fundamental four-dimensional
theory, is to compute the effective interactions among the $k_0{=}0$ modes
generated by integrating out the $k_0{\not=}0$ modes.
For instance, diagrams like
\begin {equation}
   \vcenter{
     \hbox{
        
        \epsfbox[150 150 500 700]{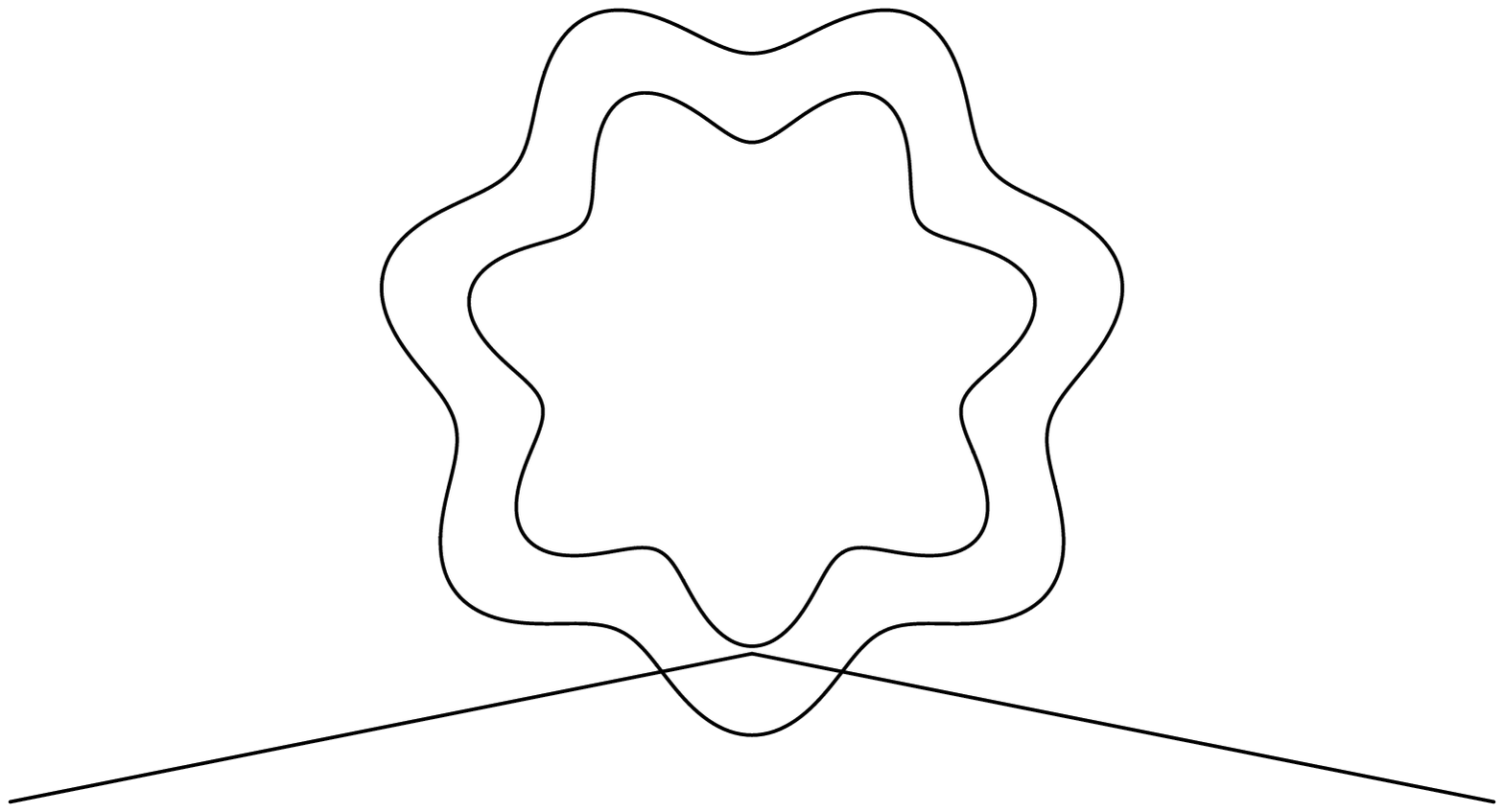}
     }
   }
   \qquad \hbox{give} \qquad
   m_3^2 = m_4^2 + g^2 T^2 ~~+~~ \hbox{higher-order} \,,
\end{equation}
where the double-lines indicate the non-static $k_0{\not=}0$ modes,
$m_4$ is the scalar mass in the fundamental four-dimensional theory,
and $m_3$ is the scalar mass in the effective three-dimensional theory.%
\footnote{
   For more details of the reduction from four to three dimensions in the
   particular context of the electroweak phase transition, see
   ref.~\citenum{Farakos}.
}
A similar thing happens for the temporal polarization of the photon.
Diagrams like
\begin {equation}
   \vcenter{
     \hbox{
        
        \epsfbox[150 300 500 500]{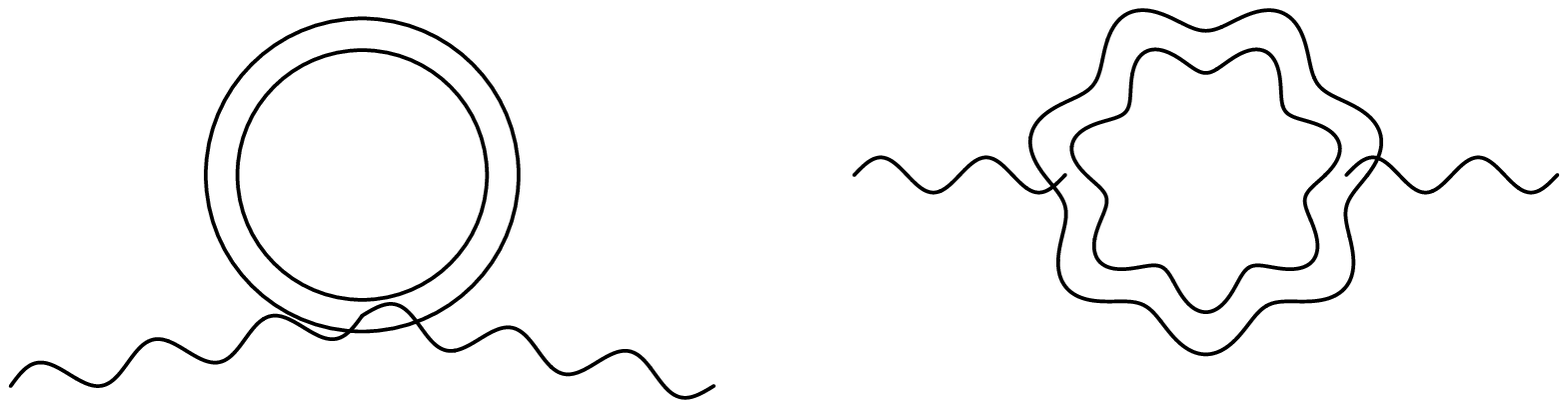}
     }
   }\
   \qquad \hbox{give} \qquad
   M_3^2(A_0) = g^2 T^2 ~~+~~ \cdots \,.
\label {debye mass}
\end{equation}
This is the Debye screening mass for static electric fields in a hot
plasma.  Coupling constants also get contributions from the $k_0{\not=}0$
modes, such as
\begin {equation}
   \vcenter{
     \hbox{
        
        \epsfbox[150 200 500 600]{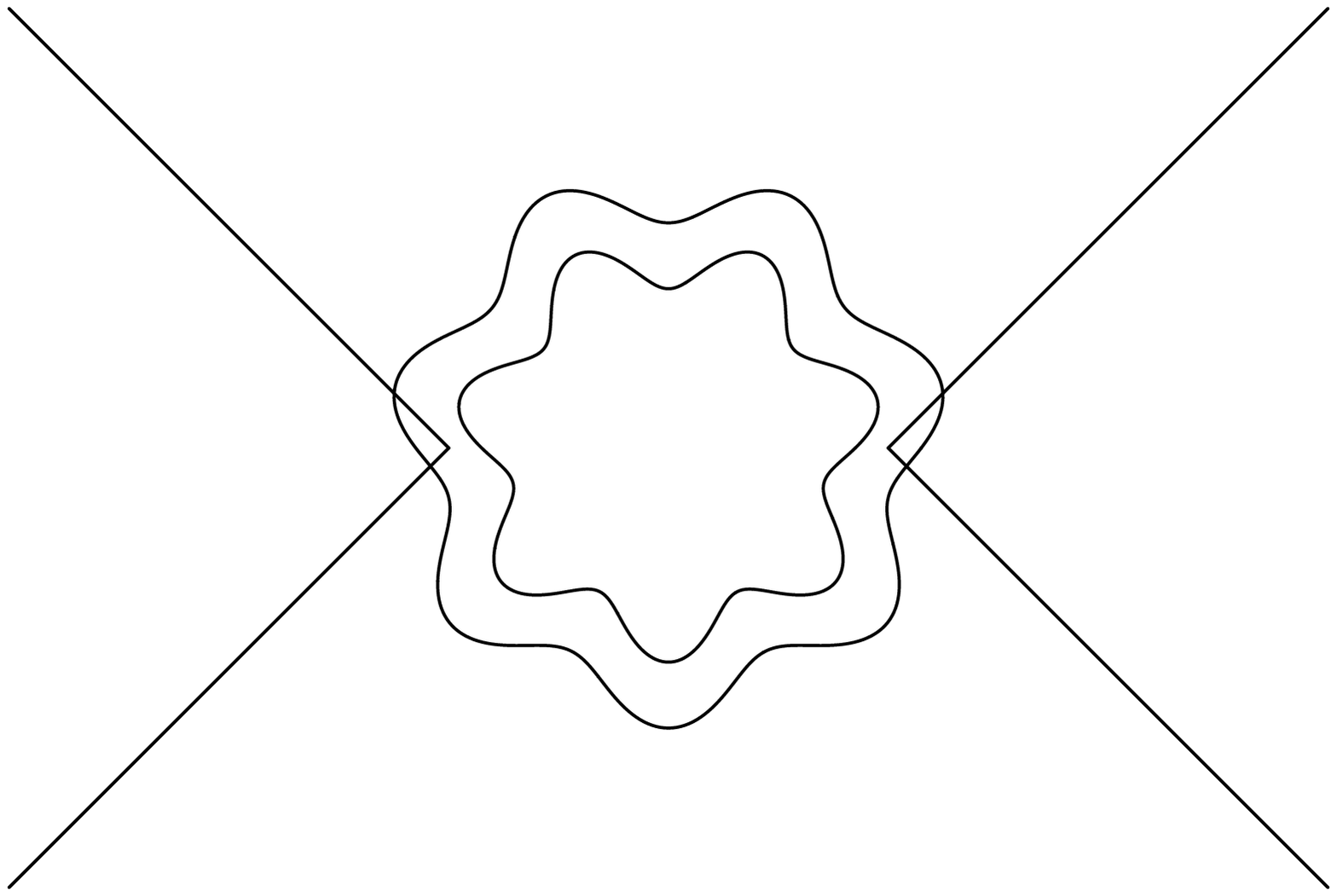}
     }
   }
   \qquad \hbox{giving} \qquad
   \lambda_3 = \lambda_4(T) ~~+~~ \cdots \,,
\end{equation}
where $\lambda_4(T)$ is the four-dimensional coupling evaluated at
renormalization scale $T$.
There is also an effective $\phi^6$ interaction, which is a marginal
operator in three dimensions:
\begin {equation}
   \vcenter{
     \hbox{
        
        \epsfbox[150 150 500 650]{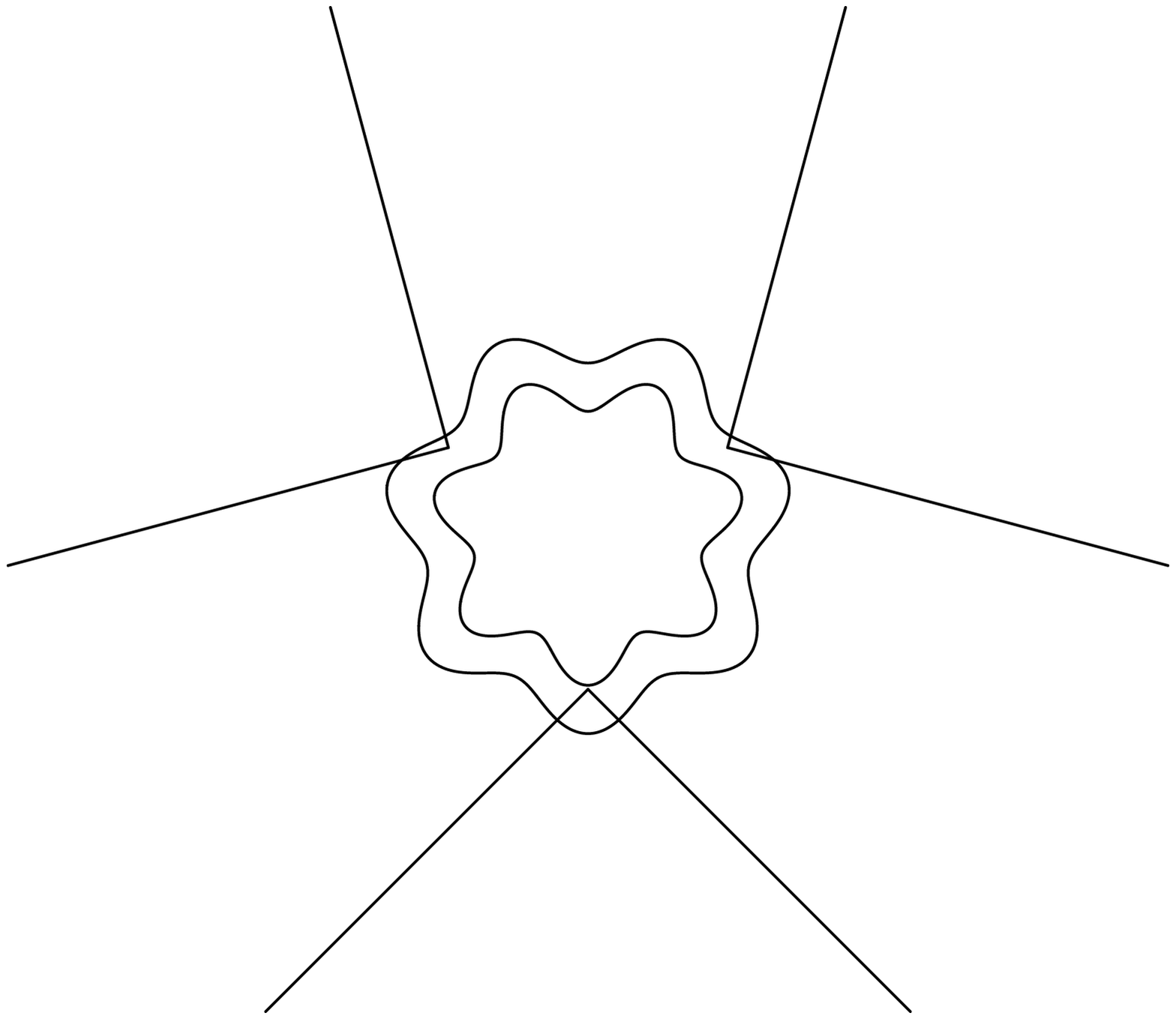}
     }
   }
   \qquad\qquad
   \hbox{marginal.}
\end {equation}
Other interactions generated by integrating out the heavy modes are
irrelevant---that is they decouple as powers of the physics scale over $T$:
\begin {equation}
   \vcenter{
     \hbox{
        
        \epsfbox[150 150 500 650]{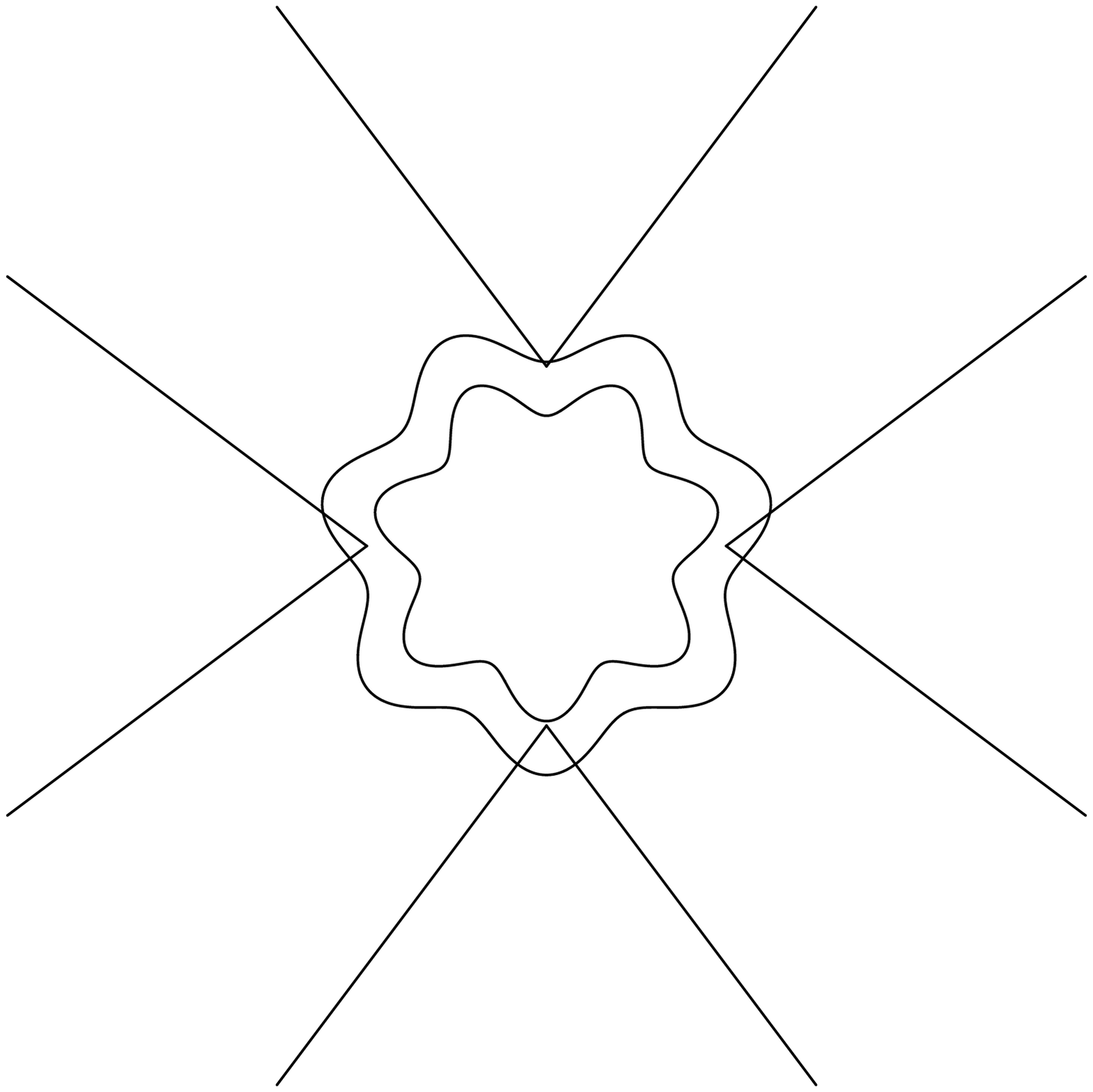}
     }
   }
   \qquad\qquad
   \hbox{irrelevant operators.}
\end {equation}

The most important point to be made about this matching of the
four-dimensional to an effective three-dimensional theory is that
the matching is {\it perturbative} in $g^2$ (and $\lambda$).
To integrate out the $k_0{=}0$ modes is to account for physics
whose scale is set by $T$ (the inverse size of the temporal dimensions)
and not by particle masses $M$.  The loop expansion parameter for this
integration is therefore
\begin {equation}
   {g^2 T \over \hbox{scale}} ~~~\sim~~~ g^2 \,,
\label {match}
\end {equation}
and everything is under the control.  It is only when one goes to
solve the effective three-dimensional theory of the $k_0{=}0$ modes
that one encounters the infrared enhancements that gave the potentially large
loop expansion parameter of section \ref{loop expansion section}.

I should also make a few remarks about choice of scale in the
regularization of the three-dimensional theory.  For simplicity, imagine
a simple UV momentum cut-off $\Lambda$.  To get the effective
three-dimensional theory, we now integrate out not only the
$k_0{\not=}0$ modes but also all $|\vec k|{>}\Lambda$
for the $k_0{=}0$ modes.  For the latter integration,
the loop expansion parameter is
\begin {equation}
   {g^2 T \over \hbox{scale}} ~~~\sim~~~ {g^2 T \over \Lambda} \,.
\end {equation}
The natural choice for $\Lambda$ is $T$---the scale at which we're doing
the matching.  If we picked $\Lambda$ to be very small, the matching
would no longer be well-behaved perturbatively.  The moral is that
the matching is simple and perturbatively calculable provided one
chooses a renormalization scale of order $T$.

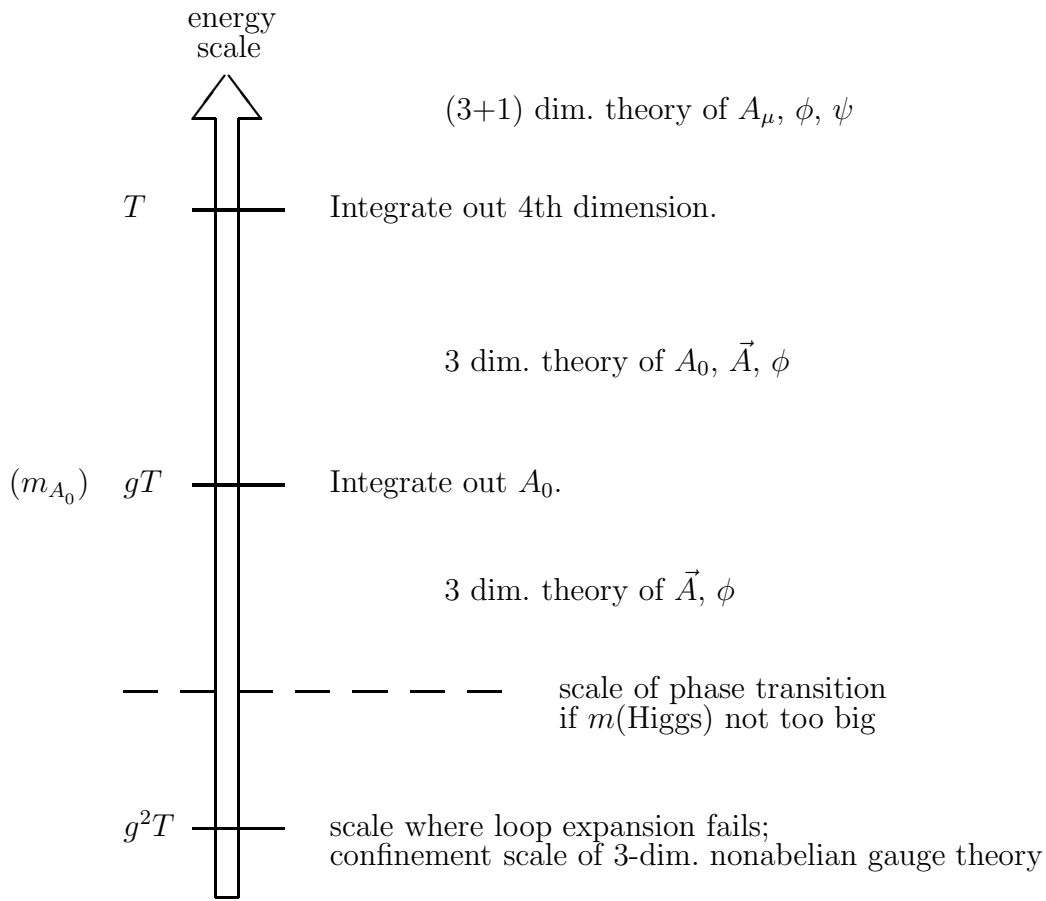
\begin{figure}
  \setlength{\unitlength}{0.12in}
  \begin {picture}(45,40)
    \thicklines
    \put(10,1){\line(1,0){1}}
    \put(10,1){\line(0,1){34}}
    \put(11,1){\line(0,1){34}}
    \put(10,35){\line(-1,0){1}}
    \put(11,35){\line( 1,0){1}}
    \put(9,35){\line( 3,4){1.4}}
    \put(12,35){\line(-3,4){1.4}}
    \put(8.8,39){energy}
    \put(9.2,37.7){scale}
    \put(9,4){\line(1,0){4}}
    \put(6,3.7){$g^2 T$}
    \put(15,3.7){scale where loop expansion fails;}
    \put(15,2.4){confinement scale of 3-dim.\ nonabelian gauge theory}
    \put(9,19){\line(1,0){4}}
    \put(6,18.7){$g T$}
    \put(1,18.7){$(m_{A_0^{}})$}
    \put(15,18.7){Integrate out $A_0$.}
    \put(9,31){\line(1,0){4}}
    \put(6,30.7){$T$}
    \put(15,30.7){Integrate out 4th dimension.}
    \put(20,35){(3+1) dim.\ theory of $A_\mu$, $\phi$, $\psi$}
    \put(20,24){3 dim.\ theory of $A_0$, $\vec A$, $\phi$}
    \put(20,14){3 dim.\ theory of $\vec A$, $\phi$}
    \put( 6  ,10){\line(1,0){1.5}}
    \put( 8.5,10){\line(1,0){1.5}}
    \put(11  ,10){\line(1,0){1.5}}
    \put(13.5,10){\line(1,0){1.5}}
    \put(16  ,10){\line(1,0){1.5}}
    \put(18.5,10){\line(1,0){1.5}}
    \put(21  ,10){\line(1,0){1.5}}
    \put(25,9.7){scale of phase transition}
    \put(25,8.4){if $m$(Higgs) not too big}
  \end {picture}
\caption {Hierarchy of scales and effective theories.  (I have shown the
  phase transition between $gT$ and $g^2T$.  If the Higgs mass is light
  enough so that $\lambda{\ll}g^3$, then it would instead be above the
  $gT$ line.)}
\label {hierarchy fig}
\end {figure}

Fig.~\ref{hierarchy fig} shows the hierarchy of important energy scales present
at finite temperature.  At scale $T$, we integrate out the fourth dimension
to get an effective 3-dimensional theory.  In the process, $A_0$ picks up
a Debye screening mass (\ref{debye mass}) of order $gT$.  So, to study
physics below $gT$, one should also integrate out $A_0$ as well.
The final effective theory at large distances is then a 3-dimensional
gauge theory of $\vec A$ and the Higgs field $\phi$.
At the scale $g^2 T$, the loop expansion parameter becomes strong and
the theory is no longer perturbatively solvable.
If $\lambda{\ll}g^2$, the scale associated with the phase transition will
be larger than $g^2 T$ and one can apply perturbative techniques.

\section {Breakdown at $\phi\sim 0$%
\protect\footnote{
   This section contains material I didn't have time to cover in my talk
   in Vladimir.
}}

\begin {figure}
\vbox
    {%
    \begin {center}
	\leavevmode
        \setlength{\unitlength}{0.15in}
        \begin {picture}(20,14)
           \put(0,0){
               
               \epsfbox [150 250 500 500] {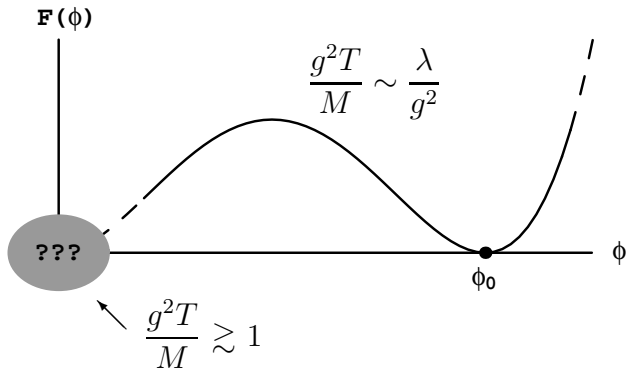}
           }
           \put(1.7,0.7){\vector(-1,1){1}}
           \put(2.3,0.1){$\displaystyle{{g^2 T\over M} \gsim 1}$}
           \put(8,9){$\displaystyle{
                {g^2 T\over M} \sim {\lambda\over g^2}    }$}
        \end {picture}
    \end {center}
    \caption
	{%
	\label {break}
        The uncertainties in $F(\phi)$ in different regions of $\phi$.
        Perturbation theory is controlled by $\lambda/g^2$ in the region
        of the hump and the asymmetric minimum, designated qualitatively
        by the solid line, but it breaks down close to the origin.
        The size of the problem region is small when $\lambda/g^2$ is
        small.
	}%
    }%
\end {figure}

Assume for the moment that $\lambda/g^2$ is arbitrarily small.
When I made the original estimate (\ref{loop parameter 2}) that the
loop expansion parameter $g^2 T/\Mw$ is of order $\lambda/g^2$, I assumed
that $\phi$ was the same order of magnitude as its value in the
asymmetric ground state.  However, since $\Mw \sim g\phi$, the loop
expansion parameter $g^2 T/\Mw$ must eventually get large as I approach
the symmetric ground-state $\phi{=}0$, no matter how small
$\lambda/g^2$ is.  This situation is depicted in fig.~\ref{break}.
For small $\lambda/g^2$, there will be a small region around
$\phi{=}0$ where perturbation theory breaks down and our calculation of
the free energy is uncertain.
How big is this uncertainty?  In particular, this uncertainty
introduces an uncertainty in computing the critical temperature
$\Tc$ at which the two ground states become degenerate, and $\Tc$
affects every other property of the transition we might compute.

There's a simple way to estimate the magnitude
of our ignorance of the free energy $F$ in the symmetric phase.%
\footnote{
  For more on this topic, try ref.~\citenum{gpy}.
}
In that
phase, we have an unbroken 3-dimensional gauge theory.  As discussed in
section \ref{loop expansion section}, the only parameter of the theory is
then $g^2 T$.  So, by dimensional analysis, the 3-dimensional action
density is then
\begin {equation}
   {1\over V} \ln Z \sim (g^2 T)^3 \,,
\end {equation}
and so
\begin {equation}
   \hbox{uncertainty in $F(0)$} ~~~\sim~~~ g^6 T^4 \,.
\end {equation}
Now compare this to the accuracy of a perturbative calculation in the
{\it asymmetric} phase, where the loop expansion parameter is small.
The uncertainty in $F(0)$ turns out to be comparable to the accuracy
of a {\it four}-loop calculation of $F$ in the asymmetric phase.
(For details on the power-counting, see
sec.~II of ref.~\citenum{Arnold&Espinosa}.)
So, in general, a perturbative treatment of the phase transition is
useful when $\lambda/g^2$ is small, but it is useful only up to a
certain order in $\lambda/g^2$.

\section {Summary}

The following are the elements of this talk that you need to remember
for Larry Yaffe's talk on the $\epsilon$ expansion.

\begin{itemize}

\item The hard part of studying the phase transition is solving a
three-dimensional theory of $\vec A$ and $\phi$.

\item That theory has a simple relationship to the original $d{=}4$
couplings if it is defined at a renormalization scale $\Lambda\sim T$:
\begin {eqnarray}
   m_3^2 &\sim& m_4^2 + g^2 T^2 \,,
\\
   g_3 &\sim& g_4(T) \,,
\\
   \lambda_3 &\sim& \lambda_4(T) \,.
\end {eqnarray}

\item The theory can be studied with straightforward perturbation theory
(at least to some order) when
\begin {equation}
   \lambda \ll g^2 \,.
\end {equation}

\end {itemize}

\end {document}